\documentclass[final,3p,times,twocolumn,authoryear]{elsarticle}

\usepackage{isotope}
\usepackage{booktabs}
\usepackage{amsmath}

\usepackage{amssymb}

\usepackage{lineno}

\journal{Applied Radiation and Isotopes}

\begin{document}

\begin{frontmatter}



\title{Simplified Efficiency Calibration Methods for Semiconductor Detectors used in Criticality Dosimetry}


\author{Victor V. Golovko}

\address{Canadian Nuclear Laboratories, 286 Plant Road, Chalk River, ON, Canada, K0J~1J0}

\begin{abstract}
{\it ``Oversimplified''} and {\it ``simplified''} methods based on true coincidence summing effect used in uncomplicated determination of the photo-peak efficiency of the semiconductor  High Purity Germanium (HPGe) detector system are suggested and verified. The methods and calibrated  $^{60}${Co} radioactive source may be used to commission any HPGe detector to use during potential criticality event. The determined accuracy of the semiconductor HPGe detector system using this method is a few percent (for the detector system used in this study it was $\simeq$8\% for oversimplified and $\simeq$5\% for simplified methods accordingly) reasonable, expected, and good enough to use for estimation of neutron dose from irradiated human blood in a potential criticality event. Moreover, if one can experimentally deduce the photo-peak efficiency for $^{60}${Co} 1333~keV $\gamma$-ray line using the suggested methods, then with a few percent accuracy this efficiency could be also used for 1369~keV $\gamma$-ray line in the decay of $^{24}${Na} isotope.
\end{abstract}




\end{frontmatter}



\section{Introduction}

A criticality accident is defined as: \\ ``An unplanned or uncontrolled nuclear excursion resulting from the assembly of a quantity of fissile material. In some cases the mass of the material may have a sufficient excess of reactivity to become prompt-critical and results in a very fast pulse of energy accompanied by a field of thermal and ionizing radiation''~(\protect\cite{RevCritAcc:2000}). Potential criticality accidents require the coordinated response of a wide range of emergency services. 
An underlying concept in the safety standards is that prevention is better than cure. Nevertheless, radiation incidents and emergencies~(\protect\cite{RevCritAcc:2000}) do occur and safety standards are necessary that define the approaches to be used in mitigating the consequences. Criticality total dose measurements are likely to be in the range from 100~mGy to 10~Gy and, because decisions have to be made quickly on the basis of the results, simple processing is also important~(\protect\cite{Thomas_2011}).

Canadian Nuclear Laboratories (CNL) are required by the  regulator agency Canadian Nuclear Safety Commission (CNSC) to maintain criticality dosimetry systems. 
Normally, Operation of Dosimetry Service is rarely the problem, but working under extreme pressure during a potential criticality accident can be stressful. If we clarify what is reasonable and expected, we can help manage our workload better, reduce our own stress,  improve productivity, and provide a clear guidance on how to increase the number of equipments that could be utilized during potential criticality accident. 

It is not possible to predict the exact numbers of injured persons in a criticality accident event because, each criticality event known to date has huge variation~(\protect\cite{RevCritAcc:2000}). As a result, there is no applicable experience to provide the insight and essential data required to formulate a detailed projection. Criticality dosimetry is expected to be in volatile, uncertain, complex and ambiguous (VUCA) conditions, where the situation will likely lead to a potentially harmful event.  Therefore, if with a moderate effort the number of semiconductor detectors that are needed during potential criticality event can be brought to a reasonable number with straightforward method, this would be helpful. 

The use of semiconductor detectors for criticality dosimetry will be provided further in details. However, for each detector system used in criticality dosimetry a calibration of the detector system is required. The common way to calibrate the detector system is to apply a sophisticated Monte Carlo calculations in order to accurately estimate efficiency of the detector system. This task is time consuming and also requires a detailed knowledge of the internal structure of detector and its composition materials. In the majority of the cases this is proprietary information of the detector manufacturer. Moreover, the Monte Carlo program quality should be tested and validated carefully (\textit{i.e.} see \cite{heikkinen2008implementation}). The quality of information coming out of Monte Carlo program cannot be better than the quality of information that went in. Therefore, when possible the measurement of the efficiency for the detector system is always much more preferable than calculations.

Measurement of blood sodium activity is usually made by measuring the gamma activity of small samples of blood. Gamma activities can be measured by a sodium iodine (\isotope{Na}\isotope{I}) crystal scintillator with a multichannel analyzer or by a high-purity germanium (HP\isotope{Ge}) detector with a multichannel analyzer.

HP\isotope{Ge} detector is utilized in this research for two reasons. First, most \isotope[23]{Na} in the blood occurs in the form of \isotope{Na}\isotope{Cl}. When irradiated by neutrons, \isotope[37]{Cl} will become \isotope[38]{Cl}, and \isotope[38]{Cl} is a gamma emitter with a radioactive half-life of 33 min. It has $\gamma$-rays of two energies, 1643 and 2167~keV, which are relatively close to $\gamma$-ray energies of \isotope[24]{Na}. Most gamma activity in blood during the first 4 h after an excursion is due to \isotope[38]{Cl}. A high-energy resolution detector is needed to differentiate between \isotope[38]{Cl} and \isotope[24]{Na} activity. The resolution of a detector is indicated by the width of its photo-peaks. A more narrow peak indicates a greater capacity of the system to distinguish between different $\gamma$-rays of closely spaced energies. Energy resolution is defined as the width of a peak halfway between its top and base. High-purity germanium systems have a typical energy resolution of a few tenths of a percentage compared to 5-10\% for \isotope{Na}\isotope{I} detectors.

Besides, HP\isotope{Ge} spectrometers used in laboratories for routine measurements are usually routinely calibrated over a broad band of $\gamma$-ray energies under a quality assurance program. Thus, these systems can be immediately utilized after accidents. However, if the energy range is outside the energy window for \isotope[24]{Na}, one needs an obvious and reliable way to calibrate detector system to be used in potential criticality event.

\section{Approach}

\begin{figure}
	\centerline{
		\includegraphics[width=\columnwidth, trim=185 163 185 170, clip=true]{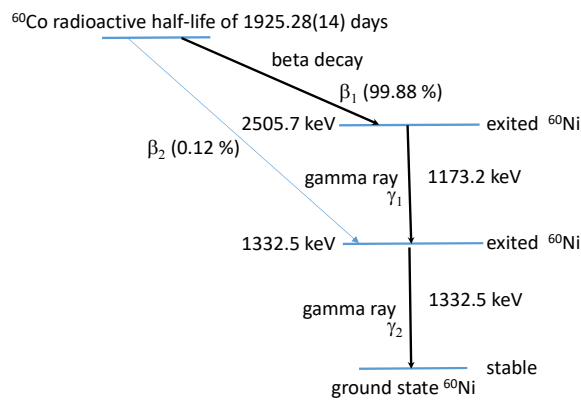}
	}
	\caption{A simplified decay scheme of a radioactive \isotope[60]{Co} nucleus (detailed decay scheme for \isotope[60]{Co} decay is shown in~\cite{TabRad_v3}). With a half-life of 5.272~years, it decays by beta-minus decay to an excited \isotope[60]{Ni} nucleus. Two $\gamma$-rays are emitted and the excitation energy is carried off, whereby the stable ground state of \isotope[60]{Ni} is reached. Provided the intermediate state (in \isotope[60]{Ni} at 1332.5~keV) is short-lived, $\gamma_1$ and $\gamma_2$ are emitted in virtual coincidence (or ``true coincidence'').}
	\label{fig:co60decay}
\end{figure}

\begin{figure}[ht]
	\centerline{
		\centering
		\includegraphics[width=\columnwidth, trim=185 163 185 170, clip=true]{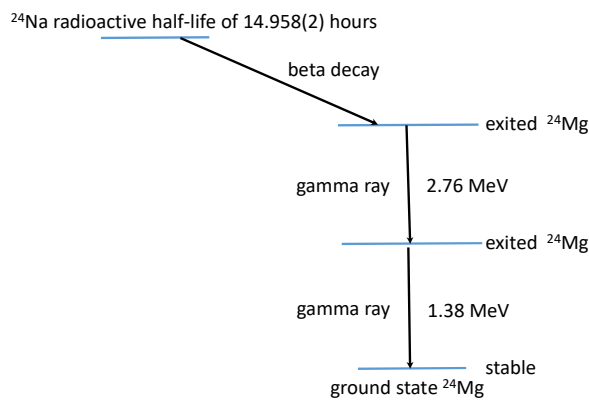} 
	}
	\caption{A simplified decay scheme of a radioactive \isotope[24]{Na} nucleus (detailed decay scheme for \isotope[24]{Na} decay is shown in~\cite{TabRad_v1}). With a half-life of 14.958~hours, it decays by beta-minus decay to an excited \isotope[24]{Mg} nucleus. Two $\gamma$-rays are rapidly emitted and the excitation energy is carried off, whereby the stable ground state of \isotope[24]{Mg} is reached.}
	\label{fig:na24decay}
\end{figure}

\isotope[60]{Co} is a synthetic radioisotope produced by the neutron activation of  \isotope[59]{Co}, which then undergoes $\beta$ decay to form \isotope[60]{Ni}, along with the emission of $\gamma$-ray radiation. \isotope[60]{Co} is extensively employed as a radiation source for calibration purpose. Namely, it is useful for criticality dosimetry application to estimate activity of  \isotope[24]{Na} in the blood as the decay scheme of \isotope[60]{Co} is similar to the decay of \isotope[24]{Na}. Moreover, 1333~keV~$\gamma$-ray energy in decay of \isotope[60]{Co} is close to 1369~keV $\gamma$-ray energy in decay of \isotope[24]{Na}. Therefore, measured HP\isotope{Ge} detector efficiency for \isotope[60]{Co} 1333~keV~$\gamma$-ray energy as a first-order of approximation could be used directly as detector efficiency for  1369~keV $\gamma$-ray energy in decay of \isotope[24]{Na}. The accuracy of the approximation used for detector calibration will be discussed further. The overall simplified decay scheme for \isotope[60]{Co} decay is shown graphically in Figure~\protect\ref{fig:co60decay} (compare it with decay of \isotope[24]{Na} shown in Figure~\protect\ref{fig:na24decay}). \\[-5pt]

\begin{figure*}[t]
	\centerline{
		\centering
		\includegraphics[width=0.99 \textwidth ]{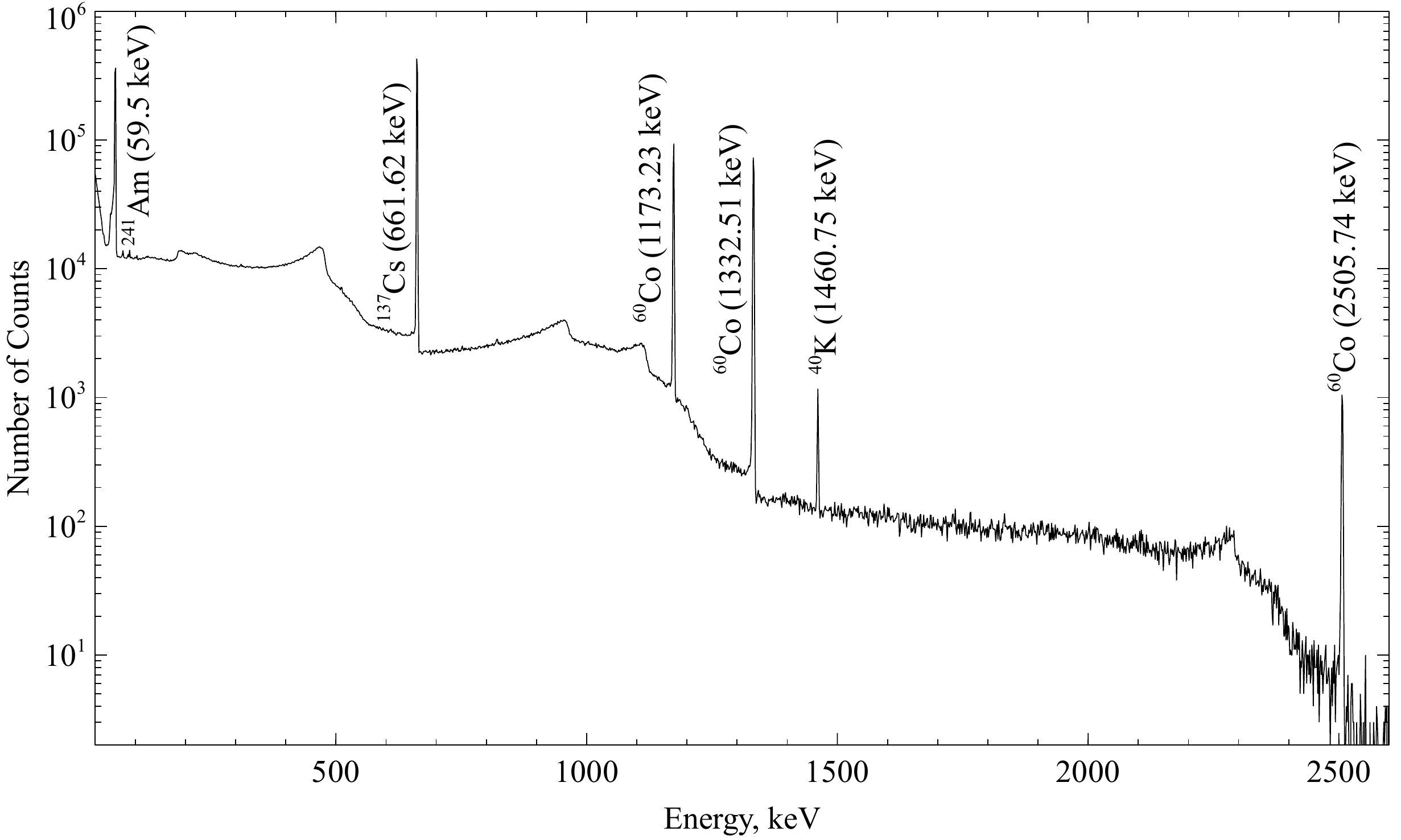} 
	}
	\caption{Measured 33.3~hours with HP\isotope{Ge} detector a calibration spectrum of the mixed radionuclide standard containing \isotope[241]{Am}, \isotope[137]{Cs}, and \isotope[60]{Co} spanning up to \isotope[60]{Co} true coincidence sum peak at 2.5~MeV. Summary of Region Of Interest (ROI) peaks shown for multinuclide calibration source shown in Table~\protect\ref{tab:roi}. }
	\label{fig:calspectrum}
\end{figure*}

The basis for nuclear spectroscopy is the location of spectral lines arising from the total absorption of charged particles or photons~(\protect\cite{Johnson:2017}). For this purpose, the resolution of the detector is important if spectral lines that are close together are to be separated and observed. HP\isotope{Ge} detectors have one of the best resolution for photon detectors. 
Nuclear spectroscopy is the analysis of radiation sources or radioisotopes by measuring the energy distribution of the source. A HP\isotope{Ge}  spectrometer is an instrument that separates the output pulses from a detector according to pulse-height (the amplitudes of electrical signals). Since the pulse-height distribution is proportional to the energy of the detected radiation, the output of the spectrometer provides detailed information that is useful to identify unknown radioisotopes and in counting one isotope in the presence of others.


The process of measuring a $\gamma$-ray begins at the radioactive source, which emits high energy photons during its unstable radioactive decay~(\protect\cite{Knoll:2010}). This spectrometry technique requires earlier learning of the photo-peak efficiency of the detector in the counting geometry for each photon energy. In order to fully utilize HP\isotope{Ge} detector as a spectrometer the energy and efficiency calibration has to be performed before using this instrument with unknown radionuclide.  

\begin{table*}[t]
\caption{ Summary of Region Of Interest (ROI) peaks shown for multi nuclide calibration source (see Figure~\protect\ref{fig:calspectrum}). The time taken for data acquisition was 120\,000 seconds or $\simeq$33.3~hours. Dead-time of detector system was maintained under 2\%.\label{tab:roi} The full width at half maximum (FWHM) is also provided for each ROI peaks.}
	{
		\resizebox{\textwidth}{!}{%
			\begin{tabular}{@{}crrrcrcrr@{}}
				\toprule
				ROI\# & \multicolumn{2}{c}{Range, keV} & Net & +/- & Centroid, keV & FWHM, keV &  Nuclide & keV \\ \midrule
				1 & 56.72 & 64.41 &  2911288 & 2122 & 60.04 & 1.30 &  \isotope[241]{Am} & 59.54 \\
				2 & 656.42 & 667.04 &  3162052 & 1867 & 661.82 & 1.71 &  \isotope[137]{Cs} & 661.62 \\
				3 & 1167.67 & 1179.03 &  635193 & 890 & 1173.41 & 2.01 &   \isotope[60]{Co} & 1173.23 \\
				4 & 1324.82 & 1339.47 & 561396 & 785 & 1332.73 & 2.09 &   \isotope[60]{Co} & 1332.51 \\
				5 & 1454.86 & 1466.95  & 6655 & 164 & 1461.12 & 2.15 &  \isotope[40]{K} & 1460.75 \\
				6 & 2497.05 & 2515.38  & 9217 & 106 & 2506.75 & 2.60 &   \isotope[60]{Co}$^a$ & 2505.74 \\ \bottomrule
			\end{tabular}%
		}
	}{$^a$This is a true coincidence sum peak in decay of \isotope[60]{Co}.
	}
\end{table*}

Detector calibrations are commonly performed for HP\isotope{Ge} detector systems. These calibrations are used by peak fitting routines to determine peak widths and estimate full-energy peak areas. Efficiency is typically described as absolute or intrinsic. The absolute efficiency is defined as the number of photons detected divided by the number of photons emitted from the source. Similarly, the intrinsic efficiency is defined as the number of photons detected divided by the number of photons incident on the detector. The intrinsic full-energy peak efficiency “depends primarily on the detector material, the incident $\gamma$-ray  energy, and the physical thickness of the detector in the direction of the incident radiation”~(\protect\cite{Nelson:2009}).

CNL External Dosimetry Service Laboratory gamma spectroscopy system efficiency calibrations are typically determined by measuring a source constructed with \\ several radionuclides of known energy and activity that span the energy range of interest in a geometry that is as close as possible to the unknown sources. For this purpose we used multinuclide calibration source containing: \isotope[241]{Am}, \isotope[137]{Cs}, and \isotope[60]{Co} that spans the energy range of interest. In this study we used HP\isotope{Ge} detector system based on Canberra GX1518 detector and multi channel analyzer Ortec DSPEC jr 2.0 with Ortec DIM-NEGGE module. Figure~\protect\ref{fig:calspectrum} shows the 33.3~hours spectrum for the multi-nuclide calibration source that spans the energy from 50~keV to 2600~keV. 

This study was performed using a certified multi nuclide standard source number 1263-23-1 produced by Eckert and Ziegler Isotope Products~(\protect\cite{isotope_product}) with a contained radioactivity of 1.012 $\mu$Ci or 37.44~kBq.\footnote{Reference date: September 1, 2007 at 12:00.} This multi nuclide standard source contained 10 radio nuclides, which are: \isotope[241]{Am}, \isotope[109]{Cd}, \isotope[57]{Co}, \isotope[123m]{Te}, \isotope[51]{Cr}, \isotope[113]{Sn}, \isotope[85]{Sr}, \isotope[137]{Cs}, \isotope[88]{Y} and \isotope[60]{Co}. Although this multi nuclide source contains many different $\gamma$-ray lines, at the time of measurements\footnote{Start of the measurement date: November 15, 2019 at 15:23.} only energy lines from \isotope[241]{Am}, \isotope[137]{Cs}, and \isotope[60]{Co} have been visible~(see Figure~\protect\ref{fig:calspectrum}). For calibration of HP\isotope{Ge} detector system source \isotope[60]{Co} with energy 1173 and 1333~keV were used with activity of 0.05486~$\mu$Ci with total uncertainty of 3.1\%.

True coincidence summing is defined as the coincident detection of photons from the same decay event, and happens for radiation from a cascade photon emitter. During decay of \isotope[60]{Co} many $\gamma$ transitions are involved. As seen in Figure~\protect\ref{fig:co60decay}, the great majority of beta decays (those labeled $\beta_1$) go to the 2505.7~keV level which falls to the ground in two steps. Thus, two $\gamma$-rays appear with their energies being the difference between the energies of the upper and lower levels: 

\begin{align*}
\gamma_1 = \left(  2505.7 - 1332.5 \right) &= 1173.2\, \rm{keV} \\
\gamma_2 = \left(  1332.5 - 0 \right) &= 1332.5\, \rm{keV}. 
\end{align*}

The two gammas are said to be in cascade, and if they appear at essentially the same time, if the intermediate level (in \isotope[60]{Ni} at 1332.5~keV) does not delay emission\footnote{The intermediate state is generally so short that the two $\gamma$-rays are, in effect, emitted in coincidence.} of the second gamma, then they are also said to be coincident (or also known as a ``true coincident'')\footnote{Another process can also lead to summed pulses due to the accidental combination of two separate events from independent decays which occur closely spaced in time. Because the time intervals separating adjacent events are randomly distributed, some will be less than the inherent resolving time of the detector or pulse processing system. These ``random coincidences'' increase rapidly with increasing counting rate, and will occur even in the absence of true coincidences.}. This phenomenon of two $\gamma$-rays appearing from the same atom at the same instant can have significant influence on counting efficiency, as will be discussed further. Additional details on true coincidence summing phenomena could be found in~(\protect\cite{ Knoll:2010,Gilmore:2011}). 

The relative number of events expected in the sum peak depends on the decay scheme, the branching ratio of the two $\gamma$-rays, the angular correlations that may exist between them, source activity, and full energy peak efficiencies, corresponding to $\gamma$-ray energies. A complete analysis is often quite complex (see for example (\protect\cite{Gilmore:2011,Debertin:1979, Debertin:1988,  Blaauw:1997, Sima:2000})), but the following simplified derivations illustrate the general approach that can be applied. For the simplified decay scheme for \isotope[60]{Co} shown in Figure~\protect\ref{fig:co60decay} we can estimate the expected number of events in the sum peak.

Let $\varepsilon_1$ be the full energy peak efficiency of the HP\isotope{Ge} detector system for gamma-ray $\gamma_1$, and $p_1$ is the photon emission intensities of gamma-ray $\gamma_1$  per disintegration. Then the full energy peak area for gamma-ray $\gamma_1$ in the absence of summing effects will be

\begin{equation}
N_1 = \varepsilon_1\,  p_1\,  A,
\label{Eq:N_1}
\end{equation}
where $A$ is the number of source decays over the observation period. Applying the same definitions to gamma-ray $\gamma_2$

\begin{equation}
N_2 = \varepsilon_2\,  p_2\,  A.
\label{Eq:N_2}
\end{equation}

The probability of simultaneous detection of both $\gamma$-rays is the product of both individual detection probabilities, multiplied by the factor $\omega(0^{\circ})$ to account for any angular correlations between the $\gamma$-ray photons. $\omega(0^{\circ})$ is defined as the relative yield of $\gamma_2$ per unit solid angle about the $0^{\circ}$ directions defined by the detector position, given that $\gamma_1$ is emitted in the same direction. Then the sum peak area should be

\begin{equation}
N_{12} = \varepsilon_1\,  p_1\, \, \varepsilon_2\,  p_2\, \, \omega(0^{\circ})\,  A.
\label{Eq:N_12}
\end{equation}

The summation process not only creates the sum peak, but also removes events which would otherwise fall within individual $\gamma$-ray full energy peaks. The remaining number of full energy events for $\gamma_1$ is (from Equation~(\protect\ref{Eq:N_1}) and~(\protect\ref{Eq:N_12}))

\begin{equation} \label{Eq:N1ws}
\begin{split}
\left . N_1 \right] {  
	\hskip -6 pt \begin{tabular}{l}
	\rm{\tiny with}  \\ [-6 pt]
	\rm{\tiny summation}
	\end{tabular}
} & = N_1 - N_{12} \\
&  = \varepsilon_1\,  p_1\,  A \left(  1- \eta_2\,  p_2 \, \omega(0^{\circ}) \right).
\end{split}
\end{equation}
To be completely accurate, in order to take into account the loss of the counts in $\gamma_1$ peak we have to take not only the photo-peak events but also losses due to other processes~(\cite{Knoll:2010}). Therefore, in Equation~(\ref{Eq:N1ws}) we changed the full energy peak efficiency $\varepsilon_2$ to the total efficiency $\eta_2$. Again, because true coincident events of any kind from $\gamma_2$ (not just photo-peak events) will remove a count from $N_1$, the detection efficiency $\eta_2$ should now be interpreted as the total efficiency. The total efficiency $\eta_2$  has either to be determined experimentally or to be calculated from the measuring geometry and the mass absorption coefficient of germanium. The first procedure is more reliable as normally the dimensions of the crystal and its distance from the front window are not well known. Applying the same approach to count the full energy peak area for gamma-ray $\gamma_2$

\begin{equation} \label{Eq:N2ws}
\begin{split}
\left . N_2 \right] {  
	\hskip -6 pt \begin{tabular}{l}
	\rm{\tiny with}  \\ [-6 pt]
	\rm{\tiny summation}
	\end{tabular}
} & = N_2 - N_{12} \\
&  = \varepsilon_2\,  p_2\,  A \left(  1- \eta_1\,  p_1 \, \omega(0^{\circ}) \right).
\end{split}
\end{equation}

The system of Equations~(\protect\ref{Eq:N1ws}), (\protect\ref{Eq:N2ws}), and (\protect\ref{Eq:N_12}) is used to experimentally determine semiconductor detector efficiency for HP\isotope{Ge} detector system to a good precision (typically 1-2\% at the 68\% confidence level between 300 and 2000~keV and 3-5\% outside of this range~(\protect\cite{GEHRKE:1977}).\footnote{The sum-peak method  has been used to determine \isotope[60]{Co} activity within 2\% deviation from the values reported by the supplier~(\protect\cite{Kim:2003}).} This system of Equations required the knowledge of the total efficiencies for two gammas in \isotope[60]{Co} decay.\footnote{The correction factors become more complicated when more than two $\gamma$-rays are emitted in cascade; however, for simplified decay of \isotope[60]{Co} we can apply this approach.} For the radio-nuclides emitting $\gamma$-rays in a simple two-step cascade in a close-geometry the efficiency (photo-peak and total) calibration method has been developed by~\protect\cite{vidmar:2003},  applied and tested for point sources of \isotope[60]{Co}, \isotope[46]{Sc} and \isotope[94]{Nb}. The same authors stressed on importance of properly taking into account angular correlation between the two $\gamma$-rays emitted in the cascade for the determination of efficiencies by applying the method to different distances of the sources from the detector.


According to the current Canadian technical and quality assurance requirements for dosimetry services regulatory document REGDOC 2.7.2~Vol.~II (\protect\cite{CNSC:2020}) the accuracy specifications for neutron dosimeters are defined as
 mean response $ 0.7 \le \overline{R} \le 1.5 $ and coefficient of variation $ cov \le 0.25 $. The response of a dosimeter, $ \overline{R} $, is defined as the result of a measurement under defined conditions, divided by the conventionally true dose that would be received under those conditions. Therefore, we would like to use a reasonable, expected, and good enough method for estimation of neutron dose from irradiated human blood in potential criticality event using HP\isotope{Ge} detector system. In order to do so, one will need a fast, easy, and robust method to estimate the efficiency of the HP\isotope{Ge} detector system. 

Simplified approximation to true coincidence summing phenomena presented in this paper and that will be explained further in details serves the purpose, and allows experimentally determine semiconductor detector efficiency for HP\isotope{Ge} detector system.  With regard to accuracy in criticality dosimetry the other~(i.e. \protect\cite{Medioni:1997}) use the next approach: ``The dosimetry system used must be capable of giving the neutron and gamma ray components of the dose with an uncertainty of less than $\pm$50\% within 48 h and less than $\pm$25\% four days later'', which is comparable to required by Canadian regulatory agency~(\protect\cite{CNSC:2020}).

In essence, in the simplified approximation the system of equations with three unknowns was formed from Equations~(\protect\ref{Eq:N_1}), (\protect\ref{Eq:N_2}), and (\protect\ref{Eq:N_12}). From now on, we will call this method as  ``{\it oversimplified\,}''. Three unknowns here are: $\varepsilon_1$, $\varepsilon_2$, and $A$.  This system of equations could be uniquely solved, and, as we used a known \isotope[60]{Co} activity from a calibrated source we can estimate the accuracy of oversimplified approximation.

The angular correlation of the $\gamma$-rays excited in the decay of \isotope[60]{Co} has been measured~(\protect\cite{Lemmer:1954}). In the absence of disturbing influences, the number of coincidences observed with an angular separation $\theta$ between the detectors is expected to be proportional to

\begin{equation} \label{Eq:omega_theta}
\omega(\theta) = 1 + \frac{1}{8} \cos^2 \theta + \frac{1}{24} \cos^4 \theta,
\end{equation}
or, for 0$^{\circ}$ degree

\begin{equation} \label{Eq:omega_0}
\omega(0^{\circ}) = 1.1667.
\end{equation}
The solution for the system of Equations~(\protect\ref{Eq:N_1}), (\protect\ref{Eq:N_2}), and (\protect\ref{Eq:N_12}) is:

\begin{align} \label{Eq:SysSol}
A &= \frac { N_1\, N_2 \, \omega(0^{\circ}) } { N_{12} },\nonumber  & \\ 
\varepsilon_1 &= \frac {N_{12}}{N_2 \, p_1 \, \omega(0^{\circ})}, & \\ 
\varepsilon_2 &= \frac {N_{12}}{N_1 \, p_2 \, \omega(0^{\circ})}. \nonumber
\end{align}
Substituting in set of Equations~\protect\ref{Eq:SysSol} the data values from Table~\protect\ref{tab:roi}, and emission probability~(\cite{TabRad_v3}) of two $\gamma$-rays in decay of \isotope[60]{Co} $p_1=0.9998$,  $p_2=0.9985$  one can get:
\begin{align} \label{Eq:SysSolDat}
A &= 45\,138\,241,\nonumber &  \\
\varepsilon_1 (1333 \, \rm{keV})  &= 0.01244, & \\
\varepsilon_2 (1173 \, \rm {kev})  &= 0.01409.\nonumber
\end{align}
As we mentioned before, $A$ is the number of source decays over the observation period, which in our case was $T_L=120\,000$~s; therefore, the source activity $ S $ will be 

\begin{equation} \label{Eq:S}
S  = 376 \, \rm{Bq}.
\end{equation}
Following the standard error analysis propagation rules~(\protect\cite{Knoll:2010, Gilmore:2011, Hughes:2010}) and taken error on the values due to counting statistics only (see Table~\protect\ref{tab:roi}) one can estimate statistical uncertainties for Equation~(\protect\ref{Eq:SysSolDat}) and (\protect\ref{Eq:S}) to be

\begin{align} \label{Eq:SysSolDatEr}
A &= 45\,138\,241\, \pm 528\,213,\nonumber & \\
\varepsilon_1 (1333 \, \rm{keV}) &= 0.01244\, \pm 0.00014, & \\
\varepsilon_2 (1173 \, \rm{keV}) &= 0.01409\, \pm 0.00016,\nonumber
\end{align}
and

\begin{equation} \label{Eq:S_err}
S =  376 \, \pm 4\, \rm{Bq}.
\end{equation}
The results uncertainties due to statistics is around $\simeq$1\%. Quoted uncertainties are given at the 68\% confidence level. 

The source activity of \isotope[60]{Co} could be estimated based on the data from certificate of calibration for standard source and following the well known radioactivity decay law

\begin{equation} \label{Eq:decay}
S(t) = S_0 \, \exp(-\lambda \, t),
\end{equation}
where $S(t)$ is the source activity at the time of the measurement, $S_0$ is the original  radioactivity of the source, $\lambda$ is the radioactive decay constant for \isotope[60]{Co}, and $t$ is the time interval. The radioactive decay constant is related to the radioactive half life for \isotope[60]{Co} via well known formulae

\begin{equation} \label{Eq:dc}
\lambda = \frac {\ln \,2}{T_{1/2}}.
\end{equation}
Combining Equations~(\protect\ref{Eq:decay}) and (\protect\ref{Eq:dc}) with data from Table~\protect\ref{tab:roi} and calibration source activity of \isotope[60]{Co} information one can get

\begin{equation} \label{Eq:dec_data_er}
S(t) =  0.01102(34)\, \mu \rm{Ci} = 408 (13)\, \rm{Bq}.
\end{equation}

Finally, one can compare results in the Equation~(\protect\ref{Eq:S_err}) and (\protect\ref{Eq:dec_data_er}) and see that two values are within 8\% of each other ($\frac { 408 - 376 } { 408 } \simeq 8\%$). Turning back to the general case, the relative systematic underestimation of the method gives $\simeq 8\%$ error. In other words, these are ``{\it oversimplified\,}'' assumptions (to justify the underestimation of the ``{\it oversimplified\,}'' method  compare Equation~(\ref{Eq:N_1}) with Equation~(\ref{Eq:N1ws})). Thus, it should be emphasized that we assign systematic uncertainties due to ``{\it oversimplified\,}'' method to be $\simeq 8\%$ (compare it with 1\% uncertainty due to counting statistic see Equation~(\protect\ref{Eq:SysSolDatEr}) and (\protect\ref{Eq:S_err})). 

At this point, the deduced accuracy of our oversimplified method is around 8\%, and one can ask a question is it possible to further improve the accuracy of the method and still apply a rather simple approach. As we mentioned above, the system of Equations~(\protect\ref{Eq:N1ws}), (\protect\ref{Eq:N2ws}), and (\protect\ref{Eq:N_12}) is used to experimentally determine semiconductor detector efficiency for HP\isotope{Ge} detector to precision of better than 1--2\%. However, it requires  to know not only the photo-peak efficiency for particular gamma-ray energy, but also the total efficiency of the detector for the same energy. This is quite a challenging and time consuming task. On the other hand, the true coincident event of any kind will remove a count from photo peak, and we can further assume that this event removal is solely due to photo-peak process events. This simplification allow to rewrite Equation~(\protect\ref{Eq:N1ws}), and (\protect\ref{Eq:N2ws})

\begin{equation} 
\label{Eq:N1sim}
\begin{split}
\left . N_1 \right] {  
	\hskip -6 pt \begin{tabular}{l}
	\rm{\tiny with}  \\ [-6 pt]
	\rm{\tiny summation}
	\end{tabular}
} & = N_1 - N_{12} \\
&  \simeq \varepsilon_1\,  p_1\,  A \left(  1- \varepsilon_2\,  p_2 \, \omega(0^{\circ}) \right),
\end{split}
\end{equation}
and

\begin{equation} 
\label{Eq:N2sim}
\begin{split}
\left . N_2 \right] {  
	\hskip -6 pt \begin{tabular}{l}
	\rm{\tiny with}  \\ [-6 pt]
	\rm{\tiny summation}
	\end{tabular}
} & = N_2 - N_{12} \\
&  \simeq \varepsilon_2\,  p_2\,  A \left(  1- \varepsilon_1\,  p_1 \, \omega(0^{\circ}) \right).
\end{split}
\end{equation}

As result, in the new simplified approximation the system of equations with three unknowns was formed from Equations~(\protect\ref{Eq:N_12}), (\protect\ref{Eq:N1sim}), and (\protect\ref{Eq:N2sim}). From now on, we will call this method as ``{\it simplified\,}'' to distinguish it from ``{\it oversimplified\,}''. Three unknowns here are: $\varepsilon_1$, $\varepsilon_2$, and $A$.  This system of equations could be uniquely solved and, as we mentioned previously, because here we used a known \isotope[60]{Co} activity from calibrated source we can estimate the accuracy of our new simplified approximation.

The solution for the system of Equations~(\protect\ref{Eq:N_12}), (\protect\ref{Eq:N1sim}), and (\protect\ref{Eq:N2sim}) is:

\begin{align} \label{Eq:SysSolSim}
A &= \frac { ( N_1 + N_{12} )\, (N_2 + N_{12}) \, \omega(0^{\circ}) } { N_{12} },\nonumber  & \\ 
\varepsilon_1 &= \frac {N_{12}}{ (N_2 + N_{12}) \, p_1 \, \omega(0^{\circ})}, & \\ 
\varepsilon_2 &= \frac {N_{12}}{ (N_1 + N_{12}) \, p_2 \, \omega(0^{\circ})}. \nonumber
\end{align}
As one can see Equation~(\protect\ref{Eq:SysSolSim}) is very similar to Equation~(\protect\ref{Eq:SysSol}). Substituting data values into Equation~(\protect\ref{Eq:SysSolSim}) and following the standard error analysis propagation rules

\begin{align} \label{Eq:SysSolDatErSim}
A &= 46\,543\,726\, \pm 526\,844 ,\nonumber & \\
\varepsilon_1 (1333 \, \rm{keV}) &= 0.01228\, \pm 0.00014, & \\
\varepsilon_2 (1173 \, \rm{keV}) &= 0.01385\, \pm 0.00016,\nonumber
\end{align}
and

\begin{equation} \label{Eq:S_errSim}
S =  389 \, \pm 4\, \rm{Bq}.
\end{equation}
Again, one can compare results in  Equation~(\protect\ref{Eq:S_errSim}) and (\protect\ref{Eq:dec_data_er}) and see that two values are within 5\% of each other ($\frac { 408 - 389 } { 408 } \simeq 5\%$). 
Turning back to the general case, the relative systematic underestimation of the method gives $\simeq 5\%$ error. In other words, these are ``{\it simplified\,}'' assumptions (to justify the underestimation of the ``{\it simplified\,}'' method  compare Equation~(\ref{Eq:N1ws}) with Equation~(\ref{Eq:N1sim})). Thus, it should be emphasize that we assign systematic uncertainties due to ``{\it simplified\,}'' method to be $\simeq 5\%$ (compare it with 1\% uncertainty due to counting statistic see Equation~(\protect\ref{Eq:SysSolDatErSim}) and (\protect\ref{Eq:S_errSim})).

A method for characterizing the shape of the efficiency curve of HP\isotope{Ge} detectors for detected $\gamma$-ray and its energy is established in~(\protect\cite{Hawari:1994, Hawari:1997}). The results indicate that for a given HP\isotope{Ge} detector the shape of the efficiency curve is indistinguishable from a straight line in the energy range extending from the energy of the lowest \isotope[46]{Sc} line at 889 keV to the highest \isotope[60]{Co} line at 1333 keV. Above 1333 keV results show a deviation from this behaviour as demonstrated by the \isotope[24]{Na} measurements . Using this method of calibration, the efficiency ratio for any two $\gamma$-ray lines in this energy range can be found to be within 0.1\%.

As we mentioned above, a method for the high-accuracy determination of $\gamma$-ray relative full-energy peak efficiencies was proposed by Hawari et al.~(\protect\cite{Hawari:1994, Hawari:1997}) in which they calibrated a HP\isotope{Ge}  semiconductor detector in the energy range from 700 to 1300 keV. In the work of \protect\cite{Ludington:2000}, they have tested the long-term stability of the calibration of the 94-cm$^3$ detector by re-counting two of the radionuclides after a period of 3 years; applied this method to a much larger, 70\% efficient 280-cm$^3$ detector; and extended the calibration to encompass the energy range from 434 to 2754~keV by using $\gamma$-rays from the radionuclide \isotope[108]{Ag} and \isotope[24]{Na}. The relative full-energy peak efficiency curve of a 280-cm$^3$  coaxial HPGe detector was determined to an accuracy of about 0.1\% over the energy range from 433 to 2754~keV from pairs of $\gamma$-ray lines whose emission probabilities are very accurately known.  The energy range from 433 to 2754 keV covers the $\gamma$-ray energies that we are interested in the decay of \isotope[24]{Na}, however, we are going to simplify their approach and apply a much more simple relation between photo-peak efficiencies and $\gamma$-ray energies.

The authors in (\protect\cite{Hawari:1994, Hawari:1997, Ludington:2000}) use common model to the efficiency curve of HP\isotope{Ge} detectors above 300~keV with a simple power law.

\begin{equation} \label{Eq:ln_eff}
\ln \left(  \frac {\varepsilon_2} {\varepsilon_1} \right) = m \, \ln \left(  \frac {E_2} {E_1} \right),
\end{equation}
where

\begin{equation} \label{Eq:m}
m = \frac { \ln (N_2/N_1)  - \ln (p_2/p_1) } { \ln (E_2/E_1) }.
\end{equation}

Therefore, if a pair of $\gamma$-rays are observed, $m$ can be determined from the $N$, $E$, and $p$ values. The approach that was utilized in \protect\cite{Hawari:1994, Hawari:1997, Ludington:2000} is to generalize Equation~(\protect\ref{Eq:ln_eff}) to a Taylor series expansion of $\ln (\varepsilon/\varepsilon_0)$  in powers of $\ln (E/E_0)$, where $E_0$ is some arbitrary reference energy, perhaps 1 or 1333~keV (the energy of the higher energy $\gamma$-ray in \isotope[60]{Co} decay) and $\varepsilon_0$  is the corresponding efficiency at that energy. The full application of the Taylor series to the Equation~(\protect\ref{Eq:ln_eff}) one could find in the original papers (see for example (\protect\cite{Ludington:2000})), here we would like to truncate the series by keeping only the first term that will simplify the final result for the photo-peak efficiency

\begin{equation} \label{Eq:ln_eff_app}
\frac {\varepsilon_2} {\varepsilon_1}  \simeq \exp \left( m \, \ln \left(  \frac {E_2} {E_1} \right) \right) .
\end{equation}

Substituting in the Equation~(\protect\ref{Eq:m}) the values from Table~\protect\ref{tab:roi} and gamma decay probabilities for \isotope[60]{Co} decay one can get

\begin{equation} \label{Eq:m_dat}
m = -0.9818 \, .
\end{equation}
Following the standard for the uncertainty propagation and taken statistic uncertainty  from Table~\protect\ref{tab:roi} one can estimate the accuracy for Equation~(\protect\ref{Eq:m_dat}) to be\footnote{Relative uncertainty on the result here is only 1.6\% and come solely from accounting statistic.}

\begin{equation} \label{Eq:m_dat_err}
m =  -0.982 \pm 0.016 \, .
\end{equation}

Calculating the efficiency for 1333~keV $\gamma$-ray line one can check applicability of the  approximation for the power law that we used in our case. This can be obtained by substituting Equation~(\protect\ref{Eq:m_dat}) to Equation~(\protect\ref{Eq:ln_eff_app}), and taken into account efficiency for 1173~keV $\gamma$-ray in decay of \isotope[60]{Co} obtained using oversimplified method ($\varepsilon_2 (1173 \, \rm{keV} = 0.01409$)  yields\footnote{Compare result with the photo-peak efficiency for 1333~keV $\gamma$-ray in decay of \isotope[60]{Co} obtained using solely oversimplified method $\varepsilon_1 (1333 \, \rm{keV})=0.01244$ (see Equation~(\protect\ref{Eq:SysSolDat})).}

\begin{equation} \label{Eq:ln_eff_app_val}
\varepsilon_1 (1333 \, \rm{keV})   = 0.01245.
\end{equation}

To estimate efficiency for 1369~kev $\gamma$-ray line in the decay of \isotope[24]{Na} one can substitute appropriate values in Equation~(\protect\ref{Eq:ln_eff_app}):

\begin{equation} \label{Eq:ln_eff_app_na24_val}%
\varepsilon (1369 \, \rm{keV})   = 0.01211.
\end{equation}
Taken into account the oversimplified true coincidence method systematic uncertainty of 8\% and following a standard error propagation rule one can estimate the uncertainty in Equation~(\protect\ref{Eq:ln_eff_app_val}) and (\protect\ref{Eq:ln_eff_app_na24_val}) to be

\begin{equation} \label{Eq:1332kev_eff_err}
\varepsilon_1 (1333 \, \rm{keV})   =  0.01245 \pm 0.00097 ,
\end{equation}
and

\begin{equation} \label{Eq:1369kev_eff_err}
\varepsilon (1369 \, \rm{keV})   =  0.01211 \pm 0.00097 .
\end{equation}

The uncertainty in Equation~(\protect\ref{Eq:1332kev_eff_err}) and (\protect\ref{Eq:1369kev_eff_err}), stemming from the described oversimplified true coincidence method itself, and the relative efficiencies results for the $\gamma$-rays with energy of 1333~keV (\isotope[60]{Co} decay line) and the $\gamma$-rays with energy of 1369~keV (\isotope[24]{Na} decay line) are the same within uncertainties (in other words $\frac{0.01245-0.01211}{0.01245} \simeq 3\% \le 8\%$ the accuracy from oversimplified method itself). Therefore, if one can experimentally deduce the photo-peak efficiency for 1333~keV $\gamma$-ray line using the described method itself, than with 8\% accuracy (coming from the oversimplified method itself) this efficiency could be also used for 1369~keV $\gamma$-ray line in decay of \isotope[24]{Na} isotope.

\section{Discussion}

This oversimplified true coincidence summing model demonstrates good agreement between field gamma spectrometry measurements and radiation  activity from calibrated \isotope[60]{Co} source. The ability to do uncomplicated efficiency calibration calculations for semiconductor germanium detector system may be of great importance in emergency response. In such a situation, one might simply make some basic measurements using uncalibrated \isotope[60]{Co} source and apply oversimplified true coincidence summing method to estimate photo-peak efficiency for 1333~keV $\gamma$-ray in decay of \isotope[60]{Co} and apply this efficiency to determine activity of \isotope[24]{Na} via measuring number of counts in 1369~keV $\gamma$-ray line.

The best solution to the problem of true coincidence summing corrections (see Equation~(\protect\ref{Eq:N_12}),  (\protect\ref{Eq:N1ws}), and (\protect\ref{Eq:N2ws})) is of course to calibrate the detector with an equally sized standard source of
the nuclide under study. In that case, coincidence summing effects need not be considered. Bearing in mind that to use a ``{\it oversimplified\,}'' (see Equation~(\protect\ref{Eq:N_1}), (\protect\ref{Eq:N_2}), and (\protect\ref{Eq:N_12})) and ``{\it simplified\,}'' (see Equations~(\protect\ref{Eq:N_12}), (\protect\ref{Eq:N1sim}), and (\protect\ref{Eq:N2sim})) true coincidence summing correction methods for calculating photo-peak efficiency of two $\gamma$-ray lines in decay of \isotope[60]{Co} one does not require the knowledge of source activity within the accuracy of the methods, and, in this study, we used this information to estimate the accuracy of ``simplification'' in the true coincidence summing method itself. Moreover, as it has been shown in previous sections the efficiency of the for 1333~keV $\gamma$-ray in decay of \isotope[60]{Co} with at least satisfactory accuracy result can be directly used as efficiency for  1369~keV $\gamma$-ray line in decay of \isotope[24]{Na}. Angular correlation effects (see Equation~(\protect\ref{Eq:N_12}) and (\protect\ref{Eq:omega_0})) cannot be neglected even in application of oversimplified true coincidence summing method.

\section{Conclusion}

This work is mainly aimed at showing that  {\it ``oversimplified''} and {\it ``simplified''} methods for coincidence summing corrections in HP\isotope{Ge} spectrometry can be calculated with an accuracy of a few percent, which is sufficient in most fields of application for potential criticality event when irradiated human blood can be used to estimate exposed neutron dose. The HP\isotope{Ge} detector system at Chalk River Laboratories Dosimetry Service can be used for criticality dosimetry. Moreover, suggested and verified methods based on true coincidence summing effect, for uncomplicated determination of the photo-peak efficiency of the semiconductor HP\isotope{Ge} detector system, is applicable for use with any HP\isotope{Ge} detector system. This method and calibrated \isotope[60]{Co} radioactive source can be used to commission any other HP\isotope{Ge} detector to use during potential criticality event. 

In addition, if one can experimentally deduce the photo-peak efficiency for \isotope[60]{Co} 1333~keV $\gamma$-ray using the suggested method, than with a few percent accuracy this efficiency could also be  used for 1369~keV $\gamma$-ray line in decay of \isotope[24]{Na} isotope. This study shows the uncomplicated and reliable way to  increase the number of detector systems for a potential criticality event, and, therefore, increases the number of potential personal neutron dose estimation during accident event in VUCA environment.

In summary, the {\it ``oversimplified''} method and {\it ``simplified''} method are well suited for the efficiency calibration of HP\isotope{Ge} detector system to use during potential criticality event. The inclusion of the angular dependence in the detector-efficiency parameter is an important part of simplified method.

\section*{Acknowledgement}

I would like to thank  everyone who made a contribution into this project. With your help we tested a simple calibration method based on true coincidence summing effect for HP\isotope{Ge} detector system. The field test used in the above mentioned approach helped in validation the current CNL criticality dosimetry process, ensure site, R\&D personnel and community safety, and fulfill our regulator agency CNSC licence requirements. 

I also owe a debt of gratitude to Cindy Hamel, Joel Vandal, and Jiansheng Sun  of CNL, for taking the data collection off regular hours in the project, preparation of HP\isotope{Ge} detector system, and all organizational steps needed to carry out this project, and the many of CNL personnel, too numerous to mention by name, for their helpful suggestions. The author wishes to thank Jiansheng Sun for his careful review of the paper.







\end{document}